%% file: ms.tex
\pdfoutput=1
\documentclass[11pt,a4paper,twoside,openright,closeany,textany]{book}
\usepackage{makeidx}
\usepackage{graphicx}
\usepackage{paralist}
\usepackage{caption} 
\usepackage{amsmath}
\usepackage{amssymb}
\usepackage{mathtools}
\usepackage{fancyhdr}
\usepackage{multicol}
\usepackage{multirow}

\usepackage{mdframed}
\usepackage[usenames,dvipsnames,svgnames,table]{xcolor}
\usepackage{array}
\usepackage[%
  colorlinks=true,%
  linkcolor=black,%
  urlcolor=black,%
  citecolor=black%
]{hyperref}
\usepackage{nameref} 
\usepackage{ifthen} 
\usepackage[font={small,sl}]{caption}
\usepackage[authoryear]{natbib}
\renewcommand\bibname{References}
\newcommand{\mychapbib}{
  \addcontentsline{toc}{section}{\bibname}
  \bibliographystyle{natbib}
  \bibliography{strucbioinf}
}
\usepackage[colorinlistoftodos]{todonotes}
\usepackage{letltxmacro}

\usepackage{tocstyle} 
\usetocstyle{standard}
\settocfeature{raggedhook}{\raggedright}
\newcommand{\bibfont}{\small\raggedright}
\def\cite{\citep}

\LetLtxMacro{\oldTodo}{\todo}
\renewcommand{\todo}[2][]{\oldTodo[#1]{TODO: #2}}

\usepackage[normalem]{ulem}

\newcommand\inwish[1]{\oldTodo[inline,color=SkyBlue]{WISH: #1}}

\newboolean{onechapter}

\newcommand{\AF}[1][~]{K.\@#1Anton#1Feenstra}
\newcommand{\SA}[1][~]{Sanne#1Abeln}

\newcommand{\HM}[1][~]{Halima#1Mouhib}

\newcommand{\BS}[1][~]{Bas#1Stringer}

\newcommand{\OI}[1][~]{Olga#1Ivanova}

\newcommand{\JG}[1][~]{\mbox{Jose}#1\mbox{Gavald\'a-Garc\'ia}}

\newcommand{\ARP}[1][~]{\mbox{Alumit}#1\mbox{Rodrigues}#1\mbox{Pereira}}

\newcommand{\orcid}[1]{\href{https://orcid.org/#1}{\raisebox{-0.7ex}{\protect\includegraphics[height=3ex]{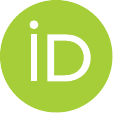}}}}
\definecolor{idgreen}{RGB}{166 206 57}
\newcommand{\mailid}[1]{\href{mailto:#1}{\raisebox{-0.3ex}{\color{idgreen}\textsf{\textbf{\Large \protect@}}}}}

\newcommand{\AFid}{\orcid{0000-0001-6755-9667}}
\newcommand{\SAid}{\orcid{0000-0002-2779-7174}}
\newcommand{\HMid}{\orcid{0000-0001-5031-3468}}
\newcommand{\JGid}{\orcid{0000-0001-6431-3442}}

\newcommand{\OIid}{\orcid{0000-0002-9111-4593}}

\newcommand{\BSid}{\orcid{0000-0001-7792-9385}}

\newcommand{\ARPid}{\mailid{alumit@gmail.com}}

\newcommand{\ACtxt}{Wrote the text}
\newcommand{\ACfig}{Created figures}
\newcommand{\ACref}{Review of current literature}
\newcommand{\ACeds}{Editorial responsibility}

\newcommand{\Angs}[1][~]{\text{\normalfont\AA}}

\renewcommand{\and}{\quad}

\newcommand{\pdbref}[1]{\href{http://www.rcsb.org/pdb/explore.do?structureId=#1}{PDB:#1}}
\newcommand{\arxiv}[2][UNDEFINED]{\href{https://arxiv.org/abs/#2}{\ifthenelse{\equal{#1}{UNDEFINED}}{arxiv.org/abs/#2}{#1}}}

\newcommand{\figref}[2][]{\hyperref[fig:#2]{Figure\@~\ref*{fig:#2}#1}}
\newcommand{\tabref}[1]{\hyperref[tab:#1]{Table \ref*{tab:#1}}}
\renewcommand{\eqref}[2][]{\hyperref[eq:#2]{Equation#1\@~\ref*{eq:#2}}}
\newcommand{\panelref}[2][]{%
    \ifthenelse{\boolean{onechapter}}{%
        \hyperref[panel:#2]{Panel\@~``\nameref{panel:#2}#1''}%
    }{%
        \hyperref[panel:#2]{Panel\@~\ref*{panel:#2}#1}%
    }%
}
\newcommand{\secref}[2][n]{%
    \hyperref[sec:#2]{%
        \ifthenelse{\equal{#1}{n} }{Section\@~\ref*{sec:#2}}{}% just number
        \ifthenelse{\equal{#1}{nn}}{Section\@~\ref*{sec:#2} ``\nameref{sec:#2}''}{}% nm & nr
        \ifthenelse{\equal{#1}{N} }{``\nameref{sec:#2}''}{}% just quoted name
        \ifthenelse{\equal{#1}{NN} }{\nameref{sec:#2}}{}% just name
    }%
}
\newcommand{\chref}[2][n]{%
    \ifthenelse{\boolean{onechapter}}{%
        \ifthenelse{\equal{#2}{ChPref}     }{\arxiv[Chapter ``\nameref*{ch:#2}'']{1801.09442}}{}%
        \ifthenelse{\equal{#2}{ChIntroPS}  }{\arxiv[Chapter ``\nameref*{ch:#2}'']{1801.09442}}{}%
        \ifthenelse{\equal{#2}{ChDetVal}   }{\arxiv[Chapter ``\nameref*{ch:#2}'']{2108.02706}}{}%
        \ifthenelse{\equal{#2}{ChStrucAli} }{\arxiv[Chapter ``\nameref*{ch:#2}'']{1801.09442}}{}%
        \ifthenelse{\equal{#2}{ChDBClass}  }{\arxiv[Chapter ``\nameref*{ch:#2}'']{1801.09442}}{}%
        \ifthenelse{\equal{#2}{ChFunc}     }{\arxiv[Chapter ``\nameref*{ch:#2}'']{1801.09442}}{}%
        \ifthenelse{\equal{#2}{ChIntroPred}}{\arxiv[Chapter ``\nameref*{ch:#2}'']{1712.00407}}{}%
        \ifthenelse{\equal{#2}{ChHomMod}   }{\arxiv[Chapter ``\nameref*{ch:#2}'']{1712.00425}}{}%
        \ifthenelse{\equal{#2}{ChSSPred}   }{\arxiv[Chapter ``\nameref*{ch:#2}'']{1801.09442}}{}%
        \ifthenelse{\equal{#2}{ChFuncPred} }{\arxiv[Chapter ``\nameref*{ch:#2}'']{1801.09442}}{}%
        \ifthenelse{\equal{#2}{ChIntroDyn} }{\arxiv[Chapter ``\nameref*{ch:#2}'']{1801.09442}}{}%
        \ifthenelse{\equal{#2}{ChThermo}   }{\arxiv[Chapter ``\nameref*{ch:#2}'']{1801.09442}}{}%
        \ifthenelse{\equal{#2}{ChMD}       }{\arxiv[Chapter ``\nameref*{ch:#2}'']{1801.09442}}{}%
        \ifthenelse{\equal{#2}{ChMC}       }{\arxiv[Chapter ``\nameref*{ch:#2}'']{1801.09442}}{}%
    }{
    \hyperref[ch:#2]{%
        \ifthenelse{\equal{#1}{n} }{Chapter \ref*{ch:#2}}{}% just number
        \ifthenelse{\equal{#1}{nn}}{Chapter \ref*{ch:#2} ``\nameref{ch:#2}''}{}% name & number
        \ifthenelse{\equal{#1}{N} }{``\nameref{ch:#2}''}{}% just name
      }%
  }%
}
\newcommand{\chrefname}[1]{\hyperref[ch:#1]{Chapter \ref*{ch:#1} ``\nameref{ch:#1}''}}
\newcommand{\partref}[1]{\hyperref[#1]{Part \ref*{#1}}}
\newcommand{\appref}[1]{\hyperref[app:#1]{Appendix \ref*{app:#1}}}

\newcommand{\figsource}[1]{\protect\footnote{Figure source location: \url{#1}}}


\newenvironment{penum}[1][\itshape i)\upshape]
{\begin{inparaenum}[#1]} {\end{inparaenum}}

\renewcommand{\arraystretch}{1.3}

\makeatletter \@beginparpenalty=5000 \makeatother

\newenvironment{bgreading}[1][]{
  \begin{mdframed}[%
      outerlinewidth=0,%
      linecolor=CornflowerBlue!30,%
      backgroundcolor=CornflowerBlue!30,%
      innerleftmargin=14,%
      innerrightmargin=14,%
    ]
	\ifthenelse{\equal{#1}{}}{}{
        \stepcounter{panel}
    	\subsection*{#1} 
    }
}{%
  \end{mdframed}
}

\usepackage{listings}
\definecolor{backcolour}{rgb}{0.95,0.95,0.92}
\definecolor{codegreen}{rgb}{0,0.6,0}
\definecolor{codegray}{rgb}{0.5,0.5,0.5}
\definecolor{codered}{rgb}{0.8,0,0.0}
\definecolor{codeblue}{rgb}{0.0,0,0.8}
\lstdefinestyle{codeStyle}{
    backgroundcolor=\color{backcolour},   
    commentstyle=\color{codegreen},
    keywordstyle=\color{codeblue},
    numberstyle=\tiny\color{codegray},
    stringstyle=\color{codegray},
    numbers=left,                    
    tabsize=2
} 
\makeindex

\begin{document}

\setboolean{onechapter}{true}

\pagestyle{fancy}
\lhead[\small\thepage]{\small\sf\nouppercase\rightmark}
\rhead[\small\sf\nouppercase\leftmark]{\small\thepage}
\newcommand{\innerfoot}{\footnotesize{\sf{\copyright} Feenstra \& Abeln}, 2014-2023}
\newcommand{\outerfoot}{\footnotesize \sf Intro Prot Struc Bioinf}
\lfoot[\outerfoot]{\innerfoot}
\cfoot{}
\rfoot[\innerfoot]{\outerfoot}
\renewcommand{\footrulewidth}{\headrulewidth}

\mainmatter
\setcounter{chapter}{3}
\chapterauthor{\JG~\JGid \and \BS~\BSid \and \OI~\OIid \and \SA*~\SAid \and \AF*~\AFid \ and \HM*~\HMid}
\chapterfigure{\includegraphics[width=0.5\linewidth]{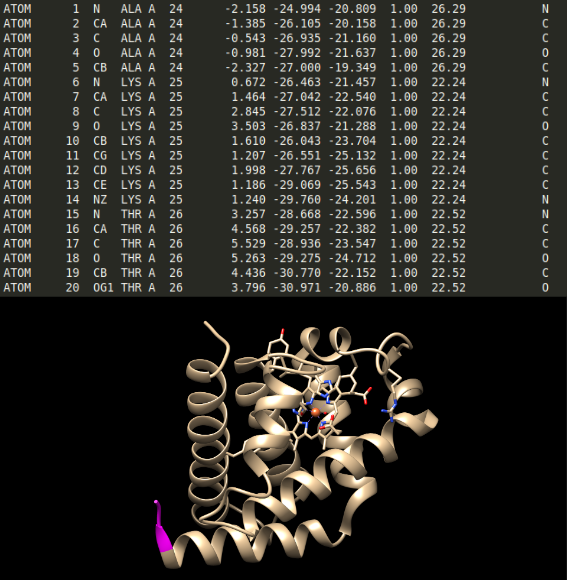}}
\chapterfootnote{* editorial responsability}
\chapter{Data Resources for Structural Bioinformatics}
\label{ch:ChDBClass}

\ifthenelse{\boolean{onechapter}}{\tableofcontents\newpage}{}

Structural bioinformatics involves a variety of computational methods, all of which require input data. Typical inputs include protein structures and sequences, which are usually retrieved from a public or private database. This chapter introduces several key resources that make such data available, as well as a handful of tools that derive additional information from experimentally determined or computationally predicted protein structures and sequences.

\section{Experimental protein structures}

Experimentally determining the structure of a protein is no easy task, as was discussed in \chref{ChDetVal}. Fortunately, researchers that do succeed in determining a protein's atomic coordinates often submit their findings to a structure database. The Protein DataBank is the largest of such databases, providing a critically important resource to structural bioinformatics research.

\subsection{The Protein DataBank}
\label{sec:ChDBClass:PDB}
\begin{figure}
\begin{center}
\includegraphics[width = 1\linewidth]{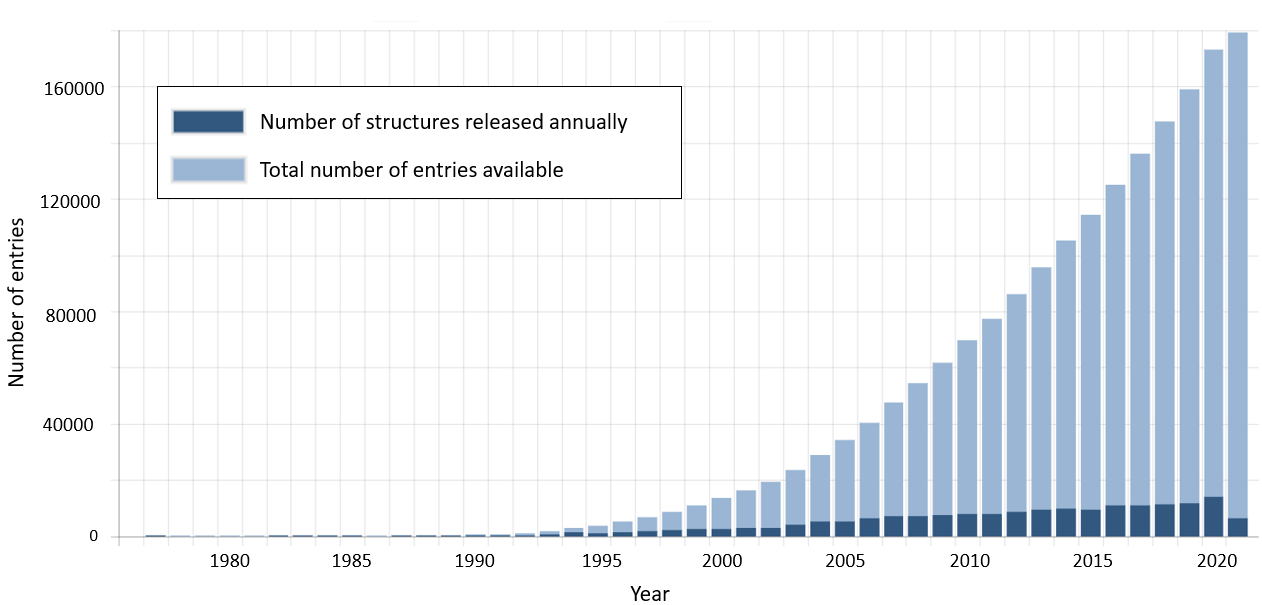}
\end{center}
\caption{The availability of protein structures in the PDB over the years (June 2021).}
\label{fig:PDBGrowth}
\end{figure}

The Protein DataBank \cite{Berman2000} (or PDB) which was established in 1971, aims to provide a freely accessible, single global archive of experimentally determined structure data for biological macromolecules. The PDB provided one of first protein crystallography structures, sperm whale myoglobin (PDB-ID: 1MBN). \figref{PDBGrowth} shows how rapidly the number of determined structures has grown over the years. It is important to realize, that the total amount of available experimental protein structures is relatively small when compared to the amount of available sequences. For example, even though the PDB contained around 173,000 structures in 2020 (\url{https://www.rcsb.org/stats/summary}), 
UniProt already contained over 180 million sequences (\url{https://www.uniprot.org/statistics/TrEMBL}).This means that for only less than 0.1\% of the known protein sequences an experimental structure of the protein is available. Note that the distribution of structures in the PDB is heavily biased towards structured proteins and proteins that are accessible for experimental structure determination. There is only little information on intrinsically disordered protein. Also, although 20-30\% of all protein sequences contain transmembrane regions, less than 2\% of protein structures in the PDB do. This is a direct consequence of experimental limitations. Since transmembrane regions usually contain large hydrophobic patches, their overexpression, purification and crystallisation is challenging and often not feasible during experimental workflows.  

\subsubsection{The PDB format}

A typical PDB entry 
contains experimental data on the 3D structure of the heavy atoms in a protein, since hydrogen atoms are difficult to detect with X-ray radiation due to their low electron density. These sets of atomic coordinates are most commonly determined through one of three different techniques: X-ray crystallography tends to resolve structures with the highest resolution, followed by Nuclear Magnetic Resonance (NMR). Cryo-electron miscroscopy (cryo-EM) is more suitable for complexes. These techniques are fundamentally different in terms of what they can and cannot measure, and how accurately they do so. Therefore, it is important to be aware of the limitations of experimental structure determination techniques when we want to use them to build computational models.

\begin{figure}[h]
\begin{center}
\includegraphics[width = 0.6\linewidth]{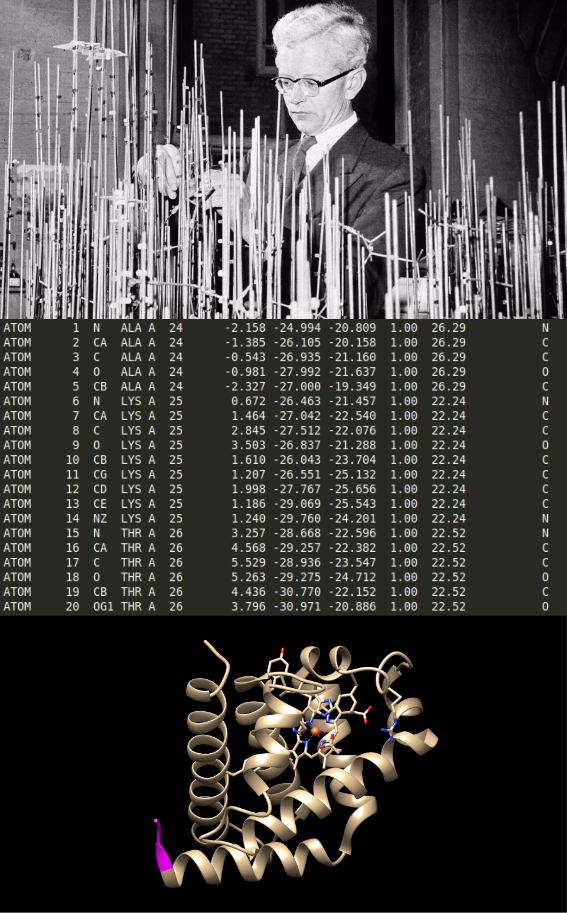}
\end{center}
\caption{Different representation of the atomic coordinates of sperm whale myoglobin (PDB-ID: 1MBN). \textbf{Top}: John Kendrew working on his atomic model. \textbf{Middle}: small section of the corresponding PDB file (see text for further explanation). \textbf{Bottom}: Cartoon representation of the protein structure using UCSF-Chimera. The residues of the first 20 atoms in the PDB file are highlighted in magenta.}
\label{fig:kendrew_struct}
\end{figure}

One of the PDB's many important contributions to structural bioinformatics, is its standardised file format. Per PDB ID you can download the protein complex in ``the PDB format'', which, among other information, contains a list of atoms, their coordinates, and which molecule or chain they are a part of. This information is stored in the lines starting with ``ATOM". 
Conceptually, this file contains all the information you need to place the atoms in a physical model, the same way John Kendrew did for sperm whale myoglobin, as shown in \figref{kendrew_struct} at the top. He elucidated its structure by means of X-ray crystallography, which granted him the Nobel Prize in Chemistry in 1962. The information allows to visualise the protein complex in more depth than the structure visualisation provided on the PDB web servers using visualization software such as UCSF ChimeraX\cite{Goddard2018}. All visualization software is compatible with PDB format files.

The PDB file format also contains important additional data including the reference where the structure was published, updates that have been made since publication. Most importantly, it also contains all relevant experimental details on the method that was used for the structure determination, such as the structural resolution and the R value; see \chref{ChDetVal} for details on the different experimental methods and parameters.

\begin{bgreading}[An ATOM line in the PDB file]
An ATOM line in a PDB file contains extensive information about a resolved atom in the crystal structure. The meaning of each column is explained in detail below. (Note that Hydrogen atoms are not present in most structures, since X-ray crystallography can not resolve them due to their low electron density.) 

\centerline{\includegraphics[width=\linewidth]{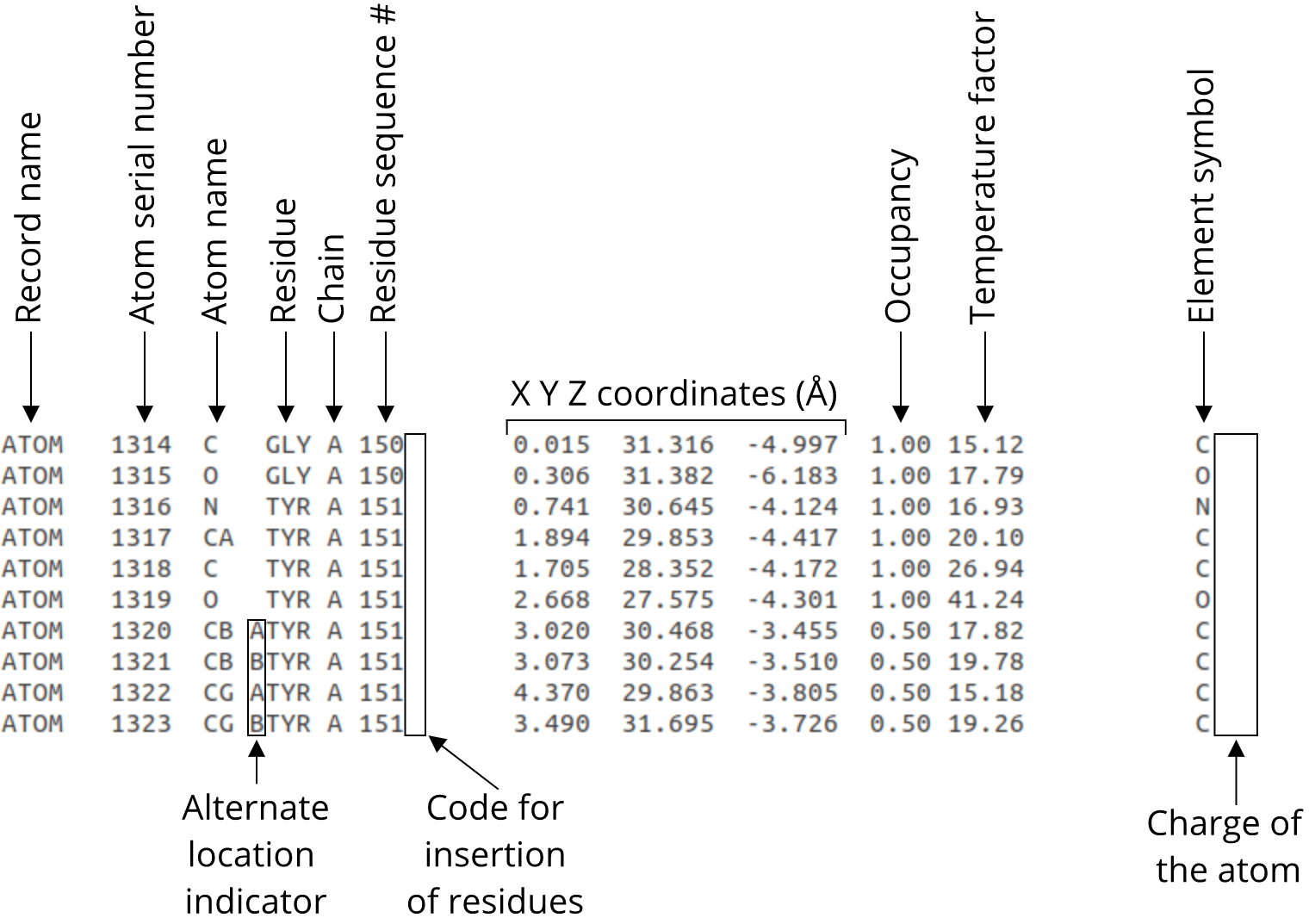}}

\begin{compactenum}
\item[\textbf{Record name}]: This field indicates the type of line in the PDB file. In this figure, all the lines are ATOM. Other types of records like REMARK, ANISOU or HETATM exist and provide different information about the protein structure
\item[\textbf{Atom serial number}]: The number of the atom in the total structural complex. Note that this number is not renewed for the next molecule in the complex. 
\item[\textbf{Atom name}]: The abbreviation of the atom name, e.g. CA is the alpha carbon. 
\item[\textbf{Alternate location indicator}]: When an atom can be found in different locations, the PDB file describes each location as a different atom. A character (A, B, C, ...) indicates to which of the different locations each entry belongs. Since each line is interpreted as an atom, it is important to note that they will have different atom serial numbers. This data is directly related to the value of occupancy. 
\item[\textbf{Residue}]: The name of the amino-acid residue to which the atom belongs, in 3-letter notation.
\item[\textbf{Chain}]: Indicates to which molecule of the structural complex this atom belongs. 
\item[\textbf{Residue sequence number}]: Indicates the position of the amino acid in the chain. Sometimes there is a ``jump" in this series, which is due to a failed elucidation of the position of the residue. This value is renewed for each chain in the structure. 
\item[\textbf{Code for insertion of residues}]: Rarely used. It is used to match the location of generally regarded ``important" amino acids in a structure, for better comparison of this structure with its different versions. For example, this could be used to ensure that a catalytic pocket of an enzyme has the same residue sequence number in structures from different isoforms or species.
\item[\textbf{X, Y and Z coordinates}]: Spatial coordinates of the atom in the X, Y and Z axis. Given in Angstroms.
\item[\textbf{Element symbol}]: Chemical symbol of the respective atom in the protein structure (C, N, S, O, H for carbo). 
\item[\textbf{Charge of the atom}]: If present, it indicates a non-neutral charge of the atom.
\item[\textbf{Occupancy}]: An occupancy of 1,0 indicates only this conformation was observed. An occupancy below 1,0 indicates that multiple conformations are possible. The occupancy of the different conformations for one atom should add up to 1.0.
\item[\textbf{Temperature factor}] or B-value: Measure of the smearing of the electron density. Higher values indicate higher excitation of the electrons and result in low precision in the determination of the atom's position.
\end{compactenum}

\end{bgreading}

\subsubsection{Details, derived data and cross references}
In addition to files like those described above, the PDB has a browser-based user interface that offers additional descriptions, derived data, and relevant cross references. For example, you can inspect a structure's Ramachandran plot (see \chref{ChIntroPS}), structure validation reports and a detailed description of the experimental methods used to determine a structure. The feature viewer displays data derived by the PDB itself (e.g.\@ secondary structure, disorder calculations, hydrophobicity) alongside information from other databases (UniProt, Pfam, Phosphosite) and homology models from the Structural Biology Knowledgebase (SBKB) and Protein Model Portal. Structural information on the PDB entries is also available on the web server PDBsum \cite{DeBeer2014}.

\begin{bgreading}[The FAIR data principles]
In recent years, there has been growing concern and awareness regarding the reusability of data. In order to improve the useful lifespan of data, more and more databases are adopting the FAIR principles \cite{Wilkinson2016}. FAIR stands for Findable, Accessible, Interoperable and Reusable, and is meant to enable the sharing of data in such a way that consumers can more effectively locate, understand and reuse it.
In the structural biology field, there was very early awareness that data should be shared in standardised formats and accompanied by ample provenance and metadata. In no small part because of this, the PDB has lived up to its vision of providing a consolidated source of experimental structure data. While this vision predates the FAIR principles by several decades, the PDB has since adopted said principles and continues to provide a wealth of structure data in a manner that suits the state of the art.
\end{bgreading}

\section{Structure analysis and annotation}

As described in \chref{ChDetVal}, experimental protein structure determination is a challenging endeavour, and computational models are often required to help and guide the process. Once protein structures are available from experiments (or from models, see \secref{ChDetVal:sequences}) an important part of structural bioinformatics comprises a full characterization of these structures by further analysing and annotating them. Topics that should be discussed in this context include structure validation, secondary structure calling, structure classification and domain definition. Many of the resources previously mentioned in this chapter actually integrate precomputed analyses and annotations in their overviews. Secondary structure, domains and related families, and structure validation reports are frequently made available.

\subsection{Structure validation}

Validating the atomic model coming out of a structure determination experiment has three aspects: confirming validity of the actual measurements; confirming if the model is consistent with said measurements; and confirming if the model respects given physical and chemical constraints. In practice, users of experimentally determined models often don't have access to the raw measurements, so the third aspect is all they can consider. For modeled structures, there are no raw measurements, so only the third aspect is relevant. 

These checks rely on tools similar to those used for homology modeling, which will be treated in depth in \chref{ChHomMod}. Structural features, such as bond lengths, angles, dihedrals, packing, H-bonds, should follow distributions similar to those observed across known (high resolution) structures. The assumption here is that these distributions will carry over to new structures, which may not hold true in case of novel features. Features that fall (far) outside the established distributions are potentially suspect, and the underlying data should be reviewed. Strong experimental evidence is required to accept an `outlier' as a bona fide observation.

A visual representation of the quality of the backbone geometry can be obtained through a Ramachandran plot (see \chref{ChIntroPS}). Strict checks  is done on whether the chemistry is in order, such as chirality of the backbone and side chains. Additional validation checks are performed using distributions which describe what is `chemically likely', and whether hydrophobic residues are inside, and hydrophilic outside. Hydrogen bonds in the (hydrophobic) protein core have to be accounted for, as even a single unsatisfied hydrogen bond donor or acceptor in a hydrophobic environment may destabilize the entire protein. Sidechain packing in the protein interior is also checked using observed distributions from known, high resolution, structures. 

There are several tools available for automated quality checking, either of novel structures or homology models. Frequently used tools include:
\begin{compactitem}
\item WhatCheck \cite{Hooft1996}
\item ProCheck \cite{Laskowski2012,Laskowski1993}
\item MolProbity \cite{Chen2010,Hintze2016}
\end{compactitem}

Secondary structure has been introduced in \chref{ChIntroPS}. Assigning secondary structures to an experimentally determined structure is a fairly straightforward exercise, and can be done manually by expert crystallographers, or with programs such as DSSP \cite{Kabsch1983} or Stride \cite{Heinig2004}. Hydrogen bonds between beta sheets, within alpha helices, or within and between other parts of a protein and surrounding molecules can easily be assigned with programs like HBplus \cite{McDonald1994}.

Note that \textit{calling} a secondary structure from atomic coordinates is a very different exercise than \textit{predicting} it from an amino acid sequence. State of the art methods, like NetSurfP-2.0 \cite{Klausen2019} and SPOT-1D \cite{Hanson2019}, can make such predictions with reasonable accuracy. For more on secondary structure prediction, see \chref{ChSSPred}.

\subsection{Structural classification}

With the set of available protein structures continuously expanding, it becomes insightful to compare and classify them. Done right, classification might for example allow us to find distant homologous relationships, since structure is generally more conserved than sequence. Well-designed structural classification schemes, such as those implemented by SCOP and CATH, can provide a gold standard for curating homologous relations. This can in turn be used to validate sequence based homology search methods, such as (PSI-)BLAST, HMMer and HHBlitz, as we discuss in \chref{ChHomMod}. 

Many features can be considered to determine whether or not two proteins should be classified together. You could for example consider sequence similarity; shared functions or functional sites; conserved secondary structure elements and topology; or a high structural alignment score, as described in \chref{ChStrucAli}. Different classification schemes may use any combination of these and other factors to group and order protein structures into a hierarchy, ideally in such a way that different levels correspond to biologically relevant features (e.g.\@ homology, or having equivalent functions).

For multi-domain proteins, structural classification works best when performed on each domain individually, rather than for the protein as a whole. Domains and domain calling are covered in the next section.

\subsubsection{SCOP}

\begin{figure}
\includegraphics[width = \linewidth]{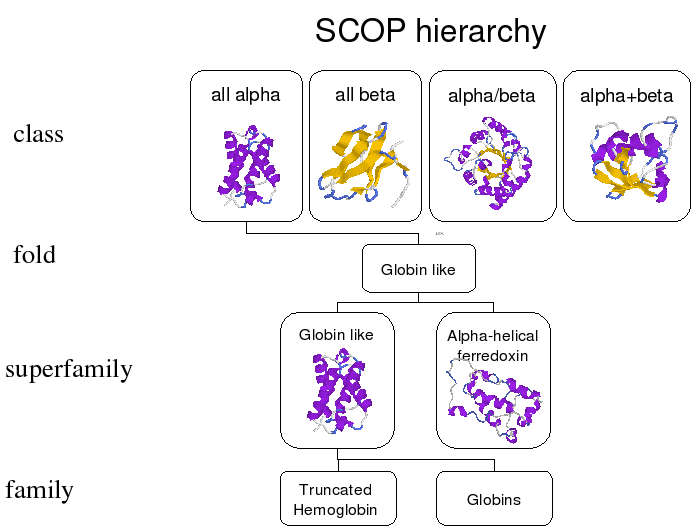}
\caption{Different levels of classification in SCOP.}
\label{fig:SCOPlevels}
\end{figure}

The Structural Classification of Proteins (SCOP) database \cite{Andreeva2008} implements four levels of hierarchy to classify protein structures. The aim is to group proteins in such a way that homologous proteins (i.e.\@ those with shared evolutionary ancestry) will cluster on the superfamily level. In order of decreasing specificity, the levels are family, superfamily, fold and class as shown in \figref{SCOPlevels}. These are manually assigned to each structure by expert curators, abetted by computational methods.

\begin{compactitem}
\item Proteins and domains are assigned to the same \emph{family} if they \begin{penum}\item have significant sequence similarity and \item have a similar function and structure\end{penum}. 

\item Families are considered part of the same \emph{superfamily} if they have low sequence identity, but their structure and functional features suggest common evolutionary origin is possible.

\item Superfamilies share a \emph{fold} if they have the same major secondary structure elements, in a similar arrangement and topology. Folds are thought to arise from certain chain topologies having specific packing arrangements that are energetically favourable.

\item \emph{Classes} are a coarse division based on the secondary structure elements: all-$\alpha$, all-$\beta$, $\alpha$/$\beta$ and $\alpha$+$\beta$, as was already introduced in \chref[nn]{ChIntroPS} (see \figref{ChIntroPS-fold-classes} for an overview). 
\end{compactitem}

\subsubsection{CATH}
				
CATH \cite{Dawson2017} is, like SCOP, a hierarchical structural classification database, based on similarity of structure, function and sequence. There are, however, some significant differences.
					
\begin{figure}
\includegraphics[width = \linewidth]{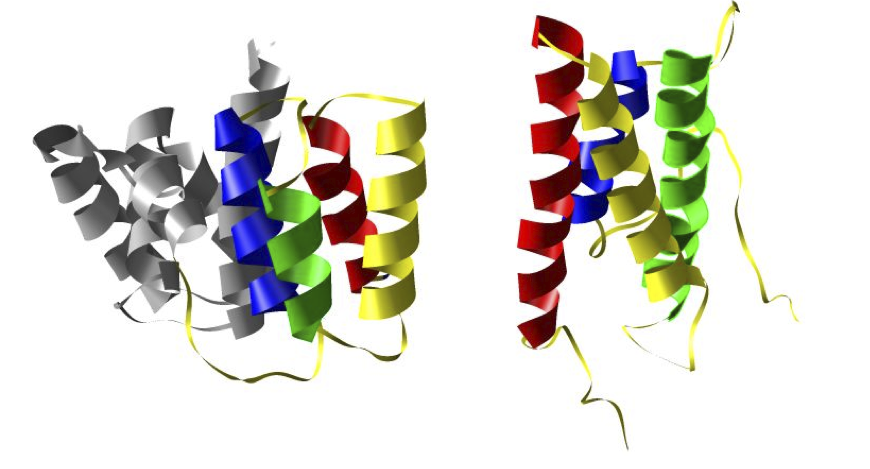}
\caption{Left: `Influenza virus matrix protein', \pdbref{1AA7}. Right: `Solution structure of four helical up-and-down bundle domain of the hypothetical protein 2610208M17Rik similar to the protein FLJ12806', \pdbref{1UG7}. CATH classifies the N-terminal domain of 1AA7 (left, coloured region) and 1UG7 (right) into the same architecture: `up and down bundle'. Following the path of the secondary structure elements (coloured sequentially: red, yellow, green and blue) it is clear that the 4 helices are differently connected and have thus another topology. SCOP classifies both proteins under the same class: `all alpha'. CATH defines two separate domains for 1AA7 (grey, coloured), whereas SCOP defines the entire protein as a single domain.}
\label{fig:Architecture}
\end{figure}

The authors of CATH attempt to automate their classification process as much as possible, without losing biological relevance. Its hierarchy consists of the levels Class, Architecture, Topology and Homology (CATH). Here topology is similar to SCOP's fold level, and homology is similar to SCOP's superfamily level. The architecture level is specific to CATH and represents the shape defined by the assembly of secondary structures without considering their connectivity (see \figref{Architecture}).
				
The biggest challenges to fully automatic assignment are recognizing domain boundaries, distinguishing between the homology and topology level, and grouping families into homologous groups. Thus, manual curation is still necessary.

\subsubsection{Domain definitions}

\begin{figure}
\centerline{\includegraphics[width = 0.9\linewidth]{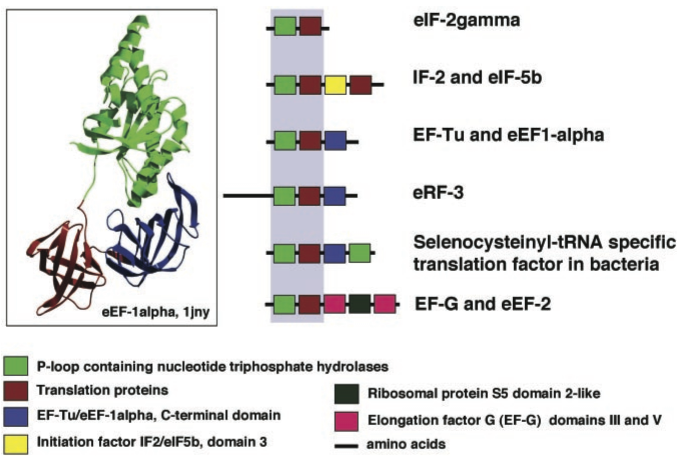}}
\caption{Domains being reused in different combinations is very common in evolution. This p-loop domain (green) occurs in at least 35 different domain combinations, six of which are shown above. From: \citet{Chothia2003}.}
\label{fig:DomainShuffling}
\end{figure}

Domains are conserved protein regions between 25 and 500 amino acids in length, which can exist and evolve independently of the rest of the protein, as already introduced briefly in \chref{ChIntroPS}. Many (but not all) domains are self-contained, and will fold into a compact structure and function independently, not unlike a single-domain protein. In evolution, it is rather common for entire domains get duplicated, deleted or inserted next to other domains like building blocks: a phenomenon referred to as domain shuffling (see \figref{DomainShuffling}, from \citet{Chothia2003}).

\begin{figure}
\includegraphics[width = \linewidth]{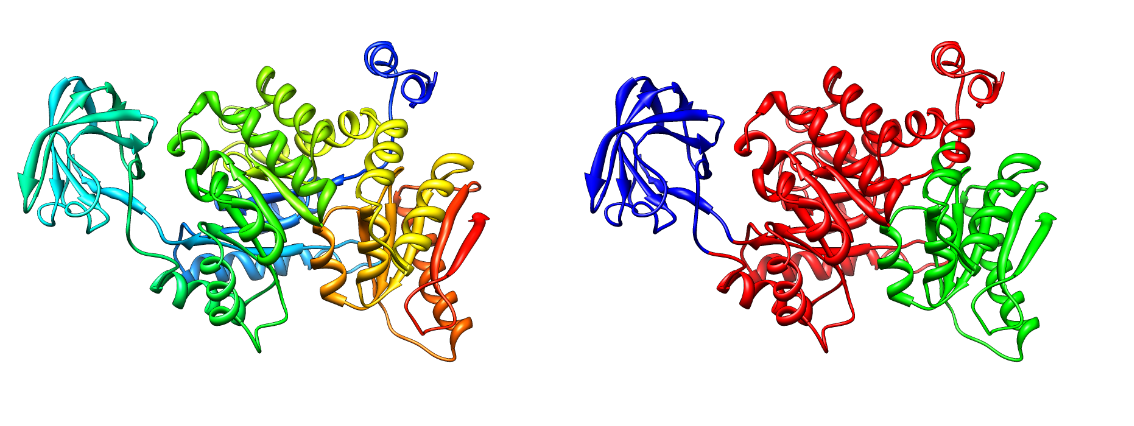}
\caption{Structure of rabbit pyruvate kinase (1PKN). Left: rainbow from blue (N) to red (C). Right: coloured according to domain boundaries as assigned by SCOP. The left domain (blue) clearly shows a distinct compact structure, with its own hydrophobic core. From this angle, the C--terminal domain (green) is not immediately apparent as a separate domain in the rainbow colored structure. Note that the middle domain (red) is discontinuous, as the blue domain sits in the middle of it.}
\label{fig:Domains1KPN}
\end{figure}

Defining domains is no trivial task. Consider for example the rainbow coloured structure in \figref{Domains1KPN}. At a glance, it is not immediately clear where the domain boundaries should fall. Several databases exist that provide domain definitions. An important distinction is domain definition that are derived from \emph{structure} and those that are derived from \emph{sequence}. We will return to domain prediction from sequence in \chref{ChIntroPred}, \secref{domains}.

Structure-based domain definition can be found in CATH, SCOP and even in the PDB. Note that the definitions between these resources do not necessarily agree. In general SCOP is more conservative in splitting up a protein in several domains; it follows the general rule that an instance of a domain, or a homolog thereof, needs to have been observed to fold independently.

Both PFAM  \cite{Finn2014} and the Conserved Domain Database (CDD) \cite{Marchler-Bauer2017} provide sequence based domain definitions. PFAM clusters protein sequences into sequence families using profile-based HMM alignments, where the seed profile has been manually curated. The CDD is a protein annotation resource that consists of domain definitions based on annotated multiple sequence alignment models. A subset of the CDD for which structures were used to define and validate domain boundaries is known as `NCBI-Curated Domains'. A position specific scoring matrix is associated with each conserved domain in the database, and can be used by PSI-BLAST for (remote) homology detection.

\subsection{Protein sequences}
\label{sec:ChDetVal:sequences}

Though not as immediately obvious as atomic coordinates, a protein's sequence contains a lot of information about its structure, and by extension, its function. Despite predicted structures becoming much more accurate and more readily available, as discussed in \chref{ChIntroPS}, there is still much more sequence data and some form of sequence analysis is an integral part of many approaches commonly used in bioinformatics research.

Due to their ubiquity, sequence databases are not as consolidated as structure databases. Many databases offer some combination of protein, RNA and DNA sequences with relevant annotations and cross references, but within the context of structural bioinformatics, UniProt is arguably the most important one.

\subsubsection{UniProt}
The UniProt database \cite{Bateman2017} consolidates a vast amount of information about proteins from a various sources. Each entry contains, at least the protein's amino acid sequence, name or description, taxonomic data, and citation information. This core data is then enriched with as many annotations as possible, among others spanning common biological ontologies, classifications and cross references, accompanied by an indication of annotation quality through experimental and computational evidence attribution. The database consist of two main parts: the UniProtKB/Swiss-Prot and the UniProtKB/TrEMBL database; resp.\@ Swiss-Prot and TrEMBL for short here. Proteins stored in TrEMBL contain protein sequences for which annotations and function characterization are created by computational techniques. As of June 2021, almost 220,000,000 protein sequences were stored in TrEMBL, although it should be noted that the protein annotations therein are not manually curated. Once a sequence in TrEMBL is reviewed the protein will be stored in Swiss-Prot which provides human curated annotations for each of the protein sequences that have been manually annotated and filtered on redundancy. In June 2021, SwissProt contained almost 560,000 curated protein sequences. 

The provided annotations in UniProt are essential to gain insight into the protein function, pathology, interaction partners, subcellular location, and post-translational modification or processing. However, not the same amount of information and annotations is provided for every sequence. The UniProt database is an essential resource for structural bioinformatics as it provides a unified framework to find sequences and refer to them by unique identifiers. Furthermore, it conveniently cross references known 3D structures, secondary structure, homologues, and family and domain annotations and provides this information through easily accessible human- and machine-readable interfaces.

\subsubsection{Modeled structure resources}

The PDB predominantly contains experimental structure data, but as mentioned before, structures are laborious and difficult to determine. Many methods and pipelines exist to predict, or model, a protein's structure in stead of direct measurements. This is discussed in greater detail in \chref{ChHomMod}.

Databases containing modeled structures include the Swiss-Model Repository (SMR) \cite{Bienert2017} and ModBase \cite{Pieper2014} databases, which offer access to structures produced by the Swiss-Model homology-modelling and ModPipe pipelines, respectively. If you are looking for several modeled structures for a particular protein of interest, the Protein Modeling Portal (PMP) \cite{Haas2013} provides a single interface to simultaneously query SMR, ModBase, and models generated by several partners of the Protein Structure Initiative (PSI) \cite{Montelione2012, MatthewZimmermanMarekGrabowski2017}.

Since the emergence of AlphaFold v2.0 by DeepMind \cite{Jumper2021}, the protein sequences in UniProt have been expanded with an unprecedented number of accompanying structural models, also known as AlphaFold DB \cite{Varadi2022}. AlphaFold generates atomic coordinates directly from the amino acid sequence. Although these AI predicted structures are extremely helpful in many cases, it is important to be aware the not all predicted structures are equally good. The predictions will be lower quality for intrinsically disordered proteins and unstructured regions of proteins. As a measure to estimate the reliability of the models, the per-residue and pairwise model-confidence estimates, as well as the `provided predicted aligned errors' should be used. 

\subsubsection{Other sequence resources}

Along with UniProt, the European Bioinformatics Institute (EMBL-EBI) also maintains UniRef \cite{Suzek2015} and UniParc \cite{Leinonen2004} -- both very useful resources as far as protein sequences are concerned, though tightly integrated with UniProt itself. In parallel, the National Center for Biotechnology Information (NCBI) maintains a plethora of databases, along with the tools to search them, including RefSeq \cite{OLeary2016}, GenBank \cite{Benson2004} and the Conserved Domain Database (CDD) \cite{Marchler-Bauer2003, Marchler-Bauer2017}. Also worth mentioning is the Protein Information Resource (PIR) \cite{Wu2003}, hosted by the Georgetown University Medical Center.

Many tools and browser interfaces exist to query and visualize the contents of these databases, such as EMBL-EBI's \textit{Ensembl} genome browser \cite{Zerbino2018}, and NCBI's \textit{Entrez} system \cite{Maglott2004}. Note that because of the high degree of cross referencing between all of the above, many search tools also yield results from, or at least refer to databases managed by other institutes.

\section{Functional data resources}

As will be further discussed in the next chapter, a protein's function is inseparably connected to its structure. Thus, an overview of important primary and tertiary structure resources would not be complete without considering correlated function annotation resources as well. Here, types of annotation to consider include, but are not limited to, protein function, interaction partners, the impacted biochemical pathways, and the location of protein expression at a cellular level. Two particularly useful examples are briefly introduced below.

\subsubsection{Function annotation: Gene Ontology}
The Gene Ontology (GO) \cite{Ashburner2000, Carbon2017} is a structured vocabulary that was designed to annotate gene products with clearly defined terms. These terms span three categories: biological process, cellular component and molecular function. Using GO terms to describe a protein's function enables searching for specific or related functions without having to deal with natural language and free text, among other benefits. As such, GO terms are very popular a annotation tool, adopted by many databases including the PDB and UniProt.

\subsubsection{Protein-protein interactions: STRING}
The Search Tool for the Retrieval of Interacting Genes/Proteins (STRING) \cite{Szklarczyk2015} is incredibly useful for compiling a list of proteins another protein of interest interacts with. STRING considers five sources of information to detect or predict your query's interaction partners: curated databases, experimental data, textmining, co-expression and homology. It additionally predicts gene neighborhoods, fusions and co-occurence, and assigns confidence scores to each interaction based on the above. STRING also allows recursive queries, to extend the resulting network with interactions to additional protein binding partners.

In summary, there exists a plethora of gene and protein annotation resources, both structural and non-structural in nature. 
Other relevant databases, that were not mention before, include the Human Phenotype Ontology (HPO) \cite{Kohler2018} describes phenotypic abnormalities in human disease; WikiPathways \cite{Slenter2018}, BioPax \cite{Demir2010} and the Kyoto Encyclopedia of Genes and Genomes (KEGG) \cite{Kanehisa2000} each contain a different type of pathway, network and model information; the Gene Expression Omnibus (GEO) \cite{Barrett2012}, Expression Atlas \cite{Petryszak2016} and Human Protein Atlas (HPA) \cite{Thul2018} track when and where certain genes are expressed.

\section{Key points}

\begin{compactitem}
\item Understand the data you are working with, and its sources
\item Experimental protein structures are available in the PDB database which provides pdb-files with atomic coordinates in 3D space.
\item Structure validation is important to ensure that a structure is based on coherent experimental data and to ensure its consistency with physical chemical constraints.
\item Databases with structural classification of proteins allow to find more distant homologues by comparison of domains and other secondary structure elements. 
\item Sequence databases are useful to find homologous proteins and their respective available information, such as a corresponding experimental or predicted structure. 
\item Other protein features can be retrieved from databases like the STRING (protein-protein interactions) or GO database (molecular function, biological process and cellular component).
\end{compactitem}

\section{Further reading}

\begin{compactitem}
\item \citet{Wilkinson2016}
\item Further explanation on FAIR principles and list of resources to expand further in each of its principles: \url{https://www.dtls.nl/fair-data/fair-principles-explained/}
\end{compactitem}

\section*{Author contributions}
{\renewcommand{\arraystretch}{1}
\begin{tabular}{@{}ll}
\ACtxt: &   JG, BS, OI, SA, KAF, HM \\
\ACfig: &   JG, BS \\
\ACref: &   JG, BS, OI, SA \\
\ACeds: &   SA, KAF, HM
\end{tabular}}

\noindent
The authors thank \ARP[ ]~\ARPid{} for critical proofreading.

\mychapbib
\clearpage

\cleardoublepage

\end{document}